\documentclass[sigconf,nonacm=true]{acmart}

\usepackage{amsmath}

\usepackage{url}
\usepackage{graphicx}
\usepackage{float}
\usepackage{xspace}
\usepackage{color}
\usepackage{balance}
\usepackage{dirtree}
\usepackage{caption}
\usepackage{subcaption}
\usepackage{comment}
\usepackage{array}
\usepackage{pifont}
\usepackage{listings}
\newcommand\note[2]{}
\newcommand\todo[1]{{\note{blue}{TODO: #1}}}

\newcommand\haris[1]{{\note{green}{haris: #1}}}

\newcommand{\etal}{\mbox{\emph{et al.\ }}}

\newcommand{\one}{({\em i})}
\newcommand{\two}{({\em ii})}

\usepackage{pifont}

\newcommand{\Onec}{{\large\ding{192}}\xspace}
\newcommand{\Twoc}{{\large\ding{193}}\xspace}
\newcommand{\Threec}{{\large\ding{194}}\xspace}
\newcommand{\Fourc}{{\large\ding{195}}\xspace}
\newcommand{\Fivec}{{\large\ding{196}}\xspace}
\newcommand{\Sixc}{{\large\ding{197}}\xspace}

\newcommand{\ovs}{{Open vSwitch}\xspace}
\newcommand{\of}{{OpenFlow}\xspace}
\newcommand{\group}{{\tt Group\_table}\xspace}
\newcommand{\httperf}{{\tt httperf}\xspace}

\newcommand{\system}{{Fractal}\xspace}
\newcommand{\System}{{Fractal}\xspace}

\begin{document}

\date{}

\title{\System: Automated Application Scaling}

\author{Masoud Koleini}
\affiliation{University of Nottingham}
\email{masoud.koleini@nottingham.ac.uk}

\author{Carlos Oviedo}
\affiliation{University of Nottingham}
\email{carlos.oviedo@nottingham.ac.uk}

\author{Derek McAuley}
\affiliation{University of Nottingham}
\email{derek.mcauley@nottingham.ac.uk}

\author{Charalampos Rotsos}
\affiliation{University of Lancaster}
\email{charalampos.rotsos@lancaster.ac.uk}

\author{Anil Madhavapeddy}
\affiliation{University of Cambridge}
\email{anil.madhavapeddy@cl.cam.ac.uk}

\author{Thomas Gazagnaire}
\affiliation{University of Cambridge}
\email{thomas.gazagnaire@cl.cam.ac.uk}

\author{Magnus Skejgstad}
\affiliation{University of Cambridge}
\email{magnus.skejgstad@cl.cam.ac.uk}

\author{Richard Mortier}
\authornote{Corresponding author. This work was originally a submission made in September 2015 to USENIX NSDI 2016.}
\affiliation{University of Cambridge}
\email{richard.mortier@cl.cam.ac.uk}

\begin{abstract}
To date, cloud applications have used datacenter resources through manual configuration and deployment of virtual machines and containers. Current trends see increasing use of \emph{microservices}, where larger applications are split into many small containers, to be developed and deployed independently. However, even with the rise of the \emph{devops} movement and orchestration facilities such as Kubernetes, there is a tendency to separate development from deployment. We present an exploration of a more extreme point on the devops spectrum: \System. Developers embed orchestration logic inside their application, fully automating the processes of scaling up and down. Providing a set of extensions to and an API over the Jitsu platform, we outline the design of \System, and describe the key features of its implementation: how an application is self-replicated, how replica lifecycles are managed, how failure recovery is handled, and how network traffic is transparently distributed between replicas. We present evaluation of a self-scaling website, and demonstrate that \System is both useful and feasible.
\end{abstract}

\maketitle
\renewcommand{\shortauthors}{Koleini et al.}

\section{Introduction}
\label{s:introduction}

\todo{The discussion in the abstract of devops and separating development from deployment does not match the rest of the paper.}

Current datacenters provide the vast pools of computing resource on which today's modern web infrastructure relies. Application deployment processes have evolved from individual hosts performing several intermingled roles, to use of virtual machines (VMs) and containers to allow hosts to support multiple cleanly separated and managed roles. To support these practices we have seen tools such as Puppet~\cite{puppet}, Ansible~\cite{ansible} and Chef~\cite{chef} developed to automate the process of building and configuring virtual machine images, and more recently, systems such as Autopilot~\cite{autopilot}, Mesos~\cite{mesos}, Borg~\cite{borg} and Kubernetes~\cite{kubernetes} to support automated management of resource use within a cluster of machines.

With improved support for orchestration of containers, many are now moving away from building monolithic web applications, to building applications as a set of loosely-coupled \emph{microservices}: small, purpose-specific web applications. All these systems explicitly separate application logic (typically wrapped up in container images) from deployment configuration (expressed as ``glue code'').


We explore an alternative approach, where both application logic and deployment configuration are written directly by the developer in the application's code, based on her insight into how her application behaves. We provide a simple API that allows her to express the conditions under which replicas of the container instance should be created. This enables more specific application-layer behaviour to be used to manage scale-up and scale-down in a deployment.~(\S\ref{s:design})

We implement this for a particular type of container on Xen, MirageOS unikernels~\cite{mirage}. Jitsu~\cite{jitsu} manages the lifecycle of replicas, and the Irmin~\cite{irmin} datastore manages state associated with each replica and, in particular, to handle merging state back when an application scales down by destroying replicas. The \ovs instance that runs on every Xen deployment, hides all replicas behind the same IP address as the first instance of the service. Although our APIs are presented specifically for MirageOS unikernels, the extensions we provide for Jitsu are generic and would enable any container or VM to make use of \System~(\S\ref{s:implementation})

We evaluate by using \System to create a website that automatically scales itself, load-balancing across replicas appropriately. We show that, as load -- defined in terms of requests/second \todo{Are we using this metric still in the last experiment?} -- increases on an instance of this website, it is able to replicate itself to maintain performance targets. Subsequently, when load decreases, the website scales down by having unloaded replicas destroy themselves, merging their website logs back into those of the first instance. We demonstrate that the overheads, in terms of memory footprint, increased response times, and OpenFlow switch rules are negligible.~(\S\ref{s:evaluation})

Load balancing and cluster resource management are long-standing areas of work, and many such systems have been developed and deployed. Most are constructed to run alongside applications, with policies concerning load metrics, thresholds and so on expressed in glue code or configuration files. In comparison, \System allows application developers to specify directly how load should be defined for their application, and how their application should behave under load.~(\S\ref{s:related})

We conclude with a brief discussion of factors that have yet to be addressed with \System. In summary, the contribution of this paper is to demonstrate the feasibility of building self-scaling applications. We do so by presenting the design and implementation of \System, a framework in which we build and evaluate a self-scaling website.~(\S\ref{s:conclusions})


\section{Design}
\label{s:design}

The key distinction between \System and existing approaches is to place control of the load balancing mechanisms directly in the application developer's hands. This enables creation of applications that adapt themselves dynamically, according to an application-specific measure of current demand and desired service. We assume that each application is hosted in a datacenter as a Xen~\cite{xen} VM, and that the \emph{dom0} control domain for each physical Xen instance runs the \ovs~\cite{ovs} softswitch to manage connectivity between local VMs. We augment this standard configuration by further assuming the VM in question is actually a MirageOS unikernel~\cite{mirage}, and that the dom0 also runs the Jitsu~\cite{jitsu} control stack; we discuss how non-MirageOS applications might use \System subsequently~(\S\ref{s:mechanics}).

Unikernels can be thought of as the combination of library operating systems with modern VMs. A unikernel VM image is created by compiling and linking application code directly with the various system libraries on which it depends. This results in compact, single-purpose VM images that boot without recourse to a traditional OS such as Linux, reducing attack surface and, in the case of MirageOS unikernels which are written in the OCaml language, leveraging modern programming language features such as type-safety and pervasive modularity~\cite{mirage}.

It is particularly notable how compact MirageOS unikernels are in comparison to traditional Linux-based VM images: the complete on-disk image for a MirageOS webserver unikernel supporting TLS is just 2\,MB, dramatically smaller than the equivalent using Apache or nginx on Linux. Jitsu is a control stack that takes advantage of this to provide extremely low-latency booting of VMs: around 30\,ms for MirageOS unikernels running on Xen/x86~\cite{jitsu}. \System provides APIs that enable a VM to request Jitsu either to boot a replica of it or to destroy it, while supporting the target VM in retaining any state it contains.

\begin{figure*}
  \centering
  \includegraphics[width=0.8\textwidth]{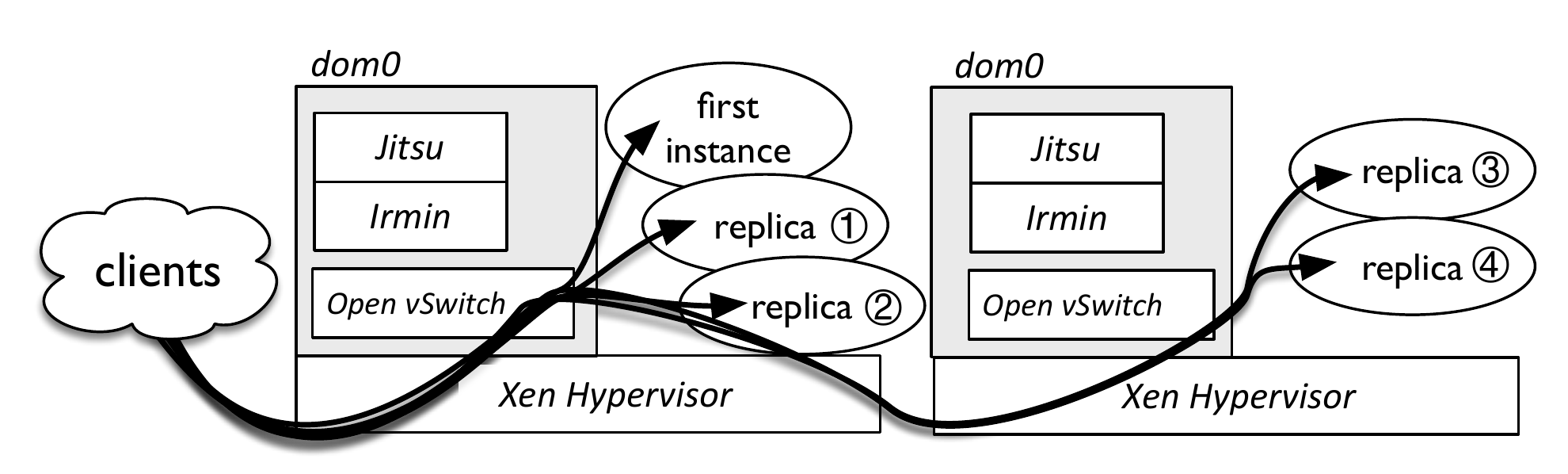}
  \caption{\label{f:topology}A \System topology for a specific application, showing the \emph{first instance}, local replicas~\Onec and~\Twoc, and remote replicas~\Threec and~\Fourc, with traffic deflected to the replicas via \ovs rules. \todo{
          The approach in Fig 1 has a bandwidth bottleneck because all inbound traffic
          must pass through the first machine's router.
      }}
\end{figure*}

Jitsu operates in a similar manner to XenStore~\cite{oxenstored}: it manages a hierarchical, transactional, key-value store. This store managed by Jitsu is an Irmin~\cite{irmin} repository, behaving similarly to a \emph{git} repository, and is shared among VMs, each VM having an allocated subtree under the parent \path{/jitsu}. Jitsu then uses the Irmin \emph{watch} mechanisms to cause events to be generated when a child of the \path{/jitsu} node is modified. Thus, by writing appropriately formatted data into their designated keys under \path{/jitsu}, a VM can trigger Jitsu to take action, creating a new replica of a VM, or self-destructing the requesting VM. Self-destruction of a VM takes place in two phases using a similar mechanism: \one~the VM indicates to Jitsu that it is halting causing Jitsu to update \ovs rules to prevent any new network flows being direct to the halting VM; and \two~when satisfactorily quiesced, the VM indicates that it should now be destroyed.

Applications that scale using \System are constrained to fit a particular application-layer topology. The first instance is booted and acts as the central node of a star. Subsequent replicas are booted following requests made to the Jitsu instance local to the requester. The detailed mechanics are described in~\S\ref{s:mechanics}, while \S~\ref{s:frecovery} discusses failure recovery both in the core and the perimenter of the star topology. Client authentication could be supported by, e.g.,~requiring client-side TLS certificates to connect to Jitsu; unfortunately, although the current release of Jitsu supports HTTPS connections using TLS, it does not yet support verification of client-side certificates. \System also relies on Irmin to enable stateful applications to manage per-replica state that must be retained after a particular replica is destroyed. Irmin provides efficient clone and merge operations, enabling git-like \emph{push} and \emph{pull} operations to be provided.
\todo{what is the access control model for cross-replica state? Single security domain for the while application?}
\todo{Given that the Fractal APIs have to interact with the datacenter policy software for allocating resources and controlling placement, doesn't each micro-service needing to interact separately make things harder, not easier, than having a pieces of dedicated deployment "glue" of existing systems?}
\todo{"The first instance is booted and acts as the central node of a star."
Oh -- if the hub crashes, this whole system stops running, yes?}
\todo{ What about health monitoring, dynamic load imbalance,
heterogeneous interacting servers, canary rollout, state management? answer: health monitoring and state management and application-specific problems, which are not required to be anwered by Fractal. }

As a running example, we will use a simple static website that monitors the number of requests/second \todo{Is this the metric we are using?} it receives, and scales up when a high threshold is exceeded and down when a low threshold is breached. This is an exceedingly simple example: this paper focuses on the overall system support required for such behaviour rather than the specific details of what metrics are best to use for a particular application -- the essence of \System is that the developer has freedom to determine her own metrics rather than being limited to traditional metrics extracted from the OS running on the VM such as CPU load, memory usage, or ping latencies or other network metrics.

\section{Implementation}
\label{s:implementation}

We next detail three key elements of our implementation: the mechanics of how a running unikernel invokes \System~(\S\ref{s:mechanics}), how we manipulate the \ovs instance to transparently deflect incoming traffic to the correct replica~(\S\ref{s:openflow}), the subsequent lifecycle management of replicas and their state~(\S\ref{s:lifecycle}), and finally, the recovery from replica and host failures~(\S\ref{s:frecovery}). The mechanisms we present make some assumptions about the datacenter environment in which they operate, but we do not believe these fundamentally limit \System{}'s applicability.
\todo{Sure, like where do you get a new VM? How do you account for limited
VM budget? What about other limited resources like bandwidth?
They don't "fundamentally limit Fractal’s applicability", but on the
other hand, these are the essential details that would help readers
understand whether this idea really works.}

\begin{figure}[H]
  \centering

  \begin{lstlisting}
type error
type response = [ `Ok | `Error of error ]
(** An invocation response. *)

type handler = response -> unit Lwt.t
(** An invocation response handler. Consumes a response and returns a schedulable thread. *)

val replicate: handler -> unit Lwt.t
(** [replicate h ()] asynchronously invokes Jitsu to boot a new copy of this unikernel, calling [h] with any response. *)

val halt: handler -> unit Lwt.t
(** [halt h ()] asynchronously invokes Jitsu to stop this unikernel receiving new flows, calling [h] with any response. *)

val die: handler -> unit Lwt.t
(** [die h ()] asynchronously invokes Jitsu to mark this unikernel as dead and ready to be garbage collected, calling [h] with any response. Note that [h] cannot be guaranteed to execute. *)
  \end{lstlisting}
  \caption{\label{f:api}\System API.}
\end{figure}

We assume that physical hosts all run XenServer, providing the Xen hypervisor and a \emph{dom0} management domain with the standard XAPI Xen management toolstack containing running instances of XenStore for storing VM configuration information and \ovs, the \of softswitch used by Xen to route traffic arriving at the physical NIC to the appropriate virtual NICs in different VMs. Each dom0 also runs an Irmin-backed Jitsu instance. The Jitsu instance is configured with any datacenter policies concerning resources available to clients, billing information and other administrative details. In the remainder of this paper we assume that each client has permission to access as many resources as are available.
\todo{Sure, like where do you get a new VM? How do you account for limited
VM budget? What about other limited resources like bandwidth?}

Every XenServer instance has its own pool of MAC addresses it can allocate to VMs at creation time. The Jitsu instance on the physical host with the IP address published for the service name launches the first instance of the application unikernel as normal, in response to a DNS request for that name. It then writes the VM settings into the Irmin store under the subtree \path{/jitsu/vms/NAME} as shown in Figure~\ref{f:irmin}. Jitsu allocates a unique identifier to each VM that it launches, and uses it as the key to the relevant subtree in Irmin and as the \group identifier for the \ovs instance on every machine hosting the unikernel's replicas (see~\S\ref{s:openflow} for details).

\subsection{Replication Mechanics}
\label{s:mechanics}


The mechanics of API invocation are straightforward. Xen imposes the constraint that only dom0 can manage VMs, and so the basic mechanism for a unikernel to create and destroy replicas is by invoking the \System API (Figure~\ref{f:api}) to write appropriate key-value pairs to the Jitsu in its dom0 to carry out the operation on its behalf via the HTTP(S) API for the Jitsu's Irmin store. Jitsu carries out the request after any necessary authentication has taken place. Jitsu follows a Unix-like access permission scheme. Each entry belongs to a specific service and its replicas have exclusive read/write access permissions. Jitsu instances has read/write access permission to its local Irmin data and to entries on remote Irmin stores that belong to hosted replicas.
\todo{again, this
reads like the paper is sweeping under the rug just the sort of details
that might make this problem challenging.}

The metadata Jitsu maintains for each unikernel is shown in Figure~\ref{f:irmin}. The first instance of each unikernel becomes the centre of a star topology, responsible for directing traffic to all replicas. This metadata includes, for all instances, the state, domain and application ids, what happens when the unikernel is stopped, the IP address of each assigned interface, and whether this is the first instance of that unikernel or not. The first instance also records the numerically lowest assigned MAC address for the unikernel, and any DNS records associated with that service. Replicas use their lowest assigned MAC address to identify the record, in place of the service name, and separately record the DNS name of the host on which the first instance is running. This allows the Jitsu responsible for booting a new replica to install the necessary \ovs rules in the first instance's dom0 to deflect traffic to the new replica appropriately.

 \begin{figure}[H]
   \centering

   \begin{lstlisting}
let poll_halt = 10 (* period to check if to halt *)
let lo_rps = 100   (* reqs/sec; low threshold *)
let hi_rps = 1000  (* reqs/sec; high threshold *)

let current_rps () =
  let rps =
    (* ...measure request per second serving... *)
  in return rps

let replicate_handler = function
  | `Ok -> (* handle successful response *)
    return ()
  | `Error e -> (* handle error *)
    return ()

let merge_state () =
  (* ...commit and merge application logs... *)
  return ()

let halt_handler = function
  | `Ok ->
    merge_state () >>= fun () ->
    die (fun _ -> return ())
  | `Error e -> return ()

let rec poll_fractal n () =
  OS.Time.sleep 1.0 >>= fun () ->
  current_rps () >>=
  fun rps ->
    if rps >= hi_rps then
      replicate replicate_handler >>= fun () ->
      poll_fractal 0 ()
    else
    if n < poll_halt then poll_fractal (n+1) ()
    else
    if rps <= lo_rps then halt halt_handler ()
    else
      poll_fractal 0 ()
 \end{lstlisting}
   \caption{\label{f:code}Simplified OCaml code extract from webserver used in~\S\ref{s:evaluation} showing \System interactions. {\tt $>>=$ } and {\tt return} are the monadic {\tt Lwt} operators that MirageOS uses for lightweight threading. }
 \end{figure}

\setlength{\DTbaselineskip}{9pt}
\begin{figure*}
  \begin{subfigure}[t]{0.95\columnwidth}
    \footnotesize
    \dirtree{%
      .1 jitsu.
      .2 vms.\DTcomment{{\em metadata for managed VMs}}.
      .3 HOSTNAME.\DTcomment{{\em service name}}.
      .4 dom-id = "16".
      .4 app-id = "0a000012".
      .4 state = "running".
      .4 stop-mode = "shutdown".
      .4 ips.
      .5 vif16.1 = "10.0.0.18".
      .5 vif16.2 = "10.0.1.18".
      .4 first-instance = "true".
      .4 mac = "12:43:3d:a3:d3:02".
      .4 dns.
      .5 service.name.
      .6 ttl = "500".
    }
    \caption{\label{f:irmin-original-vm}First instance, running on host {\tt HOSTNAME}.}
  \end{subfigure}
  \qquad%
  \begin{subfigure}[t]{0.95\columnwidth}
    \footnotesize
    \dirtree{%
      .1 jitsu.
      .2 vms.\DTcomment{{\em metadata for managed VMs}}.
      .3 MAC-ADDRESS.
      .4 dom-id = "23".
      .4 app-id = "0a000012".
      .4 state = "running".
      .4 stop-mode = "shutdown".
      .4 ip = "10.0.1.200". 
      .4 first-instance = "false".
      .4 first-instance-host = "HOSTNAME".
    }
    \caption{\label{f:irmin-remote-vm}Replica.}
  \end{subfigure}
  \caption{\label{f:irmin}Irmin key-value hierarchy used by Jitsu.}
\end{figure*}

\setlength{\DTbaselineskip}{9pt}
\begin{figure*}
  \centering
  \begin{subfigure}{0.95\columnwidth}
    \footnotesize
    \dirtree{%
      .1 jitsu.
      .2 vms.\DTcomment{{\em metadata for managed VMs}}.
      .3 REQUESTING-INSTANCE-MAC.
      .4 request = "[S(replicate); S(TARGET)]".
      .5 response = "[S(error); S(MESSAGE)]".
    }
    \caption{\label{f:irmin-local-request}Replication request from a local instance.}
  \end{subfigure}
  \qquad%
  \begin{subfigure}{0.95\columnwidth}
    \footnotesize
    \dirtree{%
      .1 jitsu.
      .2 vms.\DTcomment{{\em metadata for managed VMs}}.
      .3 HOSTNAME\DTcomment{{\em of the requesting Jitsu's host}}.
      .4 request = "[S(replicate); S(TARGET); -~I(0a000013); I(300); S(shutdown); -~S(IMAGE.xen)]".
      .5 response = "[S(success);]".
    }
    \caption{\label{f:irmin-remote-request}Replication request from a remote Jitsu to host {\tt TARGET} indicating the Xen {\tt IMAGE} to boot.}
  \end{subfigure}
  \caption{\label{f:irmin-request}Irmin key-value hierarchy used for replication request/response.}
\end{figure*}

Each Jitsu's Irmin store also records outstanding RPC invocations to create, halt, or destroy replicas. Jitsu uses Irmin's \path{watch_tag} interface over \path{/jitsu} to monitor updates to the request keys whenever they are added or updated. In addition, every Jitsu instance has a request and a response key on the Irmin stores of all other hosts in the datacenter. When a Jitsu instance runs, it starts watching its response keys on the other hosts using the \path{watch_key} interface. That helps in prompt notification of the Jitsu when it expects response from other Jitsu instances.

When a VM wishes to make an invocation, it writes the appropriate value under its \path{/jitsu/requests/INSTANCE/request} key. When the request is satisfied, the result code is written under \path{/jitsu/requests/INSTANCE/response}, which is watched by the requesting instance to determine whether the request is satisfied or not. Only one outstanding request is permitted, but the system remains responsive under load spikes due to the extremely short boot times achievable when using Jitsu to manage VMs~\cite{jitsu}.

Finally, although the API presented in Figure~\ref{f:api} is MirageOS specific the basic functionality is actually available to any VM that wishes to exercise it. All a VM must do is write appropriate key-value pairs into its local Jitsu's Irmin store. Implementation of the present API could easily be ported to other languages or shell scripts for use by any VM or container.

\subsection{Managing Traffic}
\label{s:openflow}

To load balance traffic requires consistent selection of a destination replica on a per-flow basis, with control over the proportion of traffic directed to each replica. We investigate two options for leveraging \ovs to distribute traffic appropriately.The first makes use of a non-standard primitive provided by \ovs with a two table forwarding pipeline. The first table contains exact match rules which forward traffic to a specific replica, plus a low priority default rule that uses the {\tt bundle\_load} command to hash the packet header, store an output port in a packet register, and forward all unmatched packets to the second table. The second table then contains a single {\tt learn} command, which use the outcome of the hashing function to install an exact match flow rule in the first table and forward the packet appropriately.

The second uses the standard \of~\group abstraction, introduced in \of v1.1, which aggregates forwarding operations in a \group entry, with each group having a type and identifying an ordered list of \emph{buckets}, sets of actions (i.e.,~packet modifications and forwarding) to carry out. This approach uses a single {\tt SELECT} \group entry for each replicated service; which selects at random a bucket and applies its actions on the packet. In \ovs, each set of actions in an entry can have a weight associated with it, used to influence the hashing of new flows between available group buckets.

Traffic for each application is then load-balanced among replicas using a single \group entry installed in the \ovs running in the dom0 of the original instance. Creation of a replica causes a bucket addition to the group entry that appropriately rewrites header fields (primarily IP and Ethernet destination addresses) and forwards traffic to the switch port to which the destination replica is attached. When a replica is removed, the weight of the corresponding bucket is set to zero initially so that no new flows are directed to the dying replica and the bucket entry is finally removed when all outstanding connections are completed.

\begin{figure*}
  \centering
  \includegraphics[width=.95\textwidth]{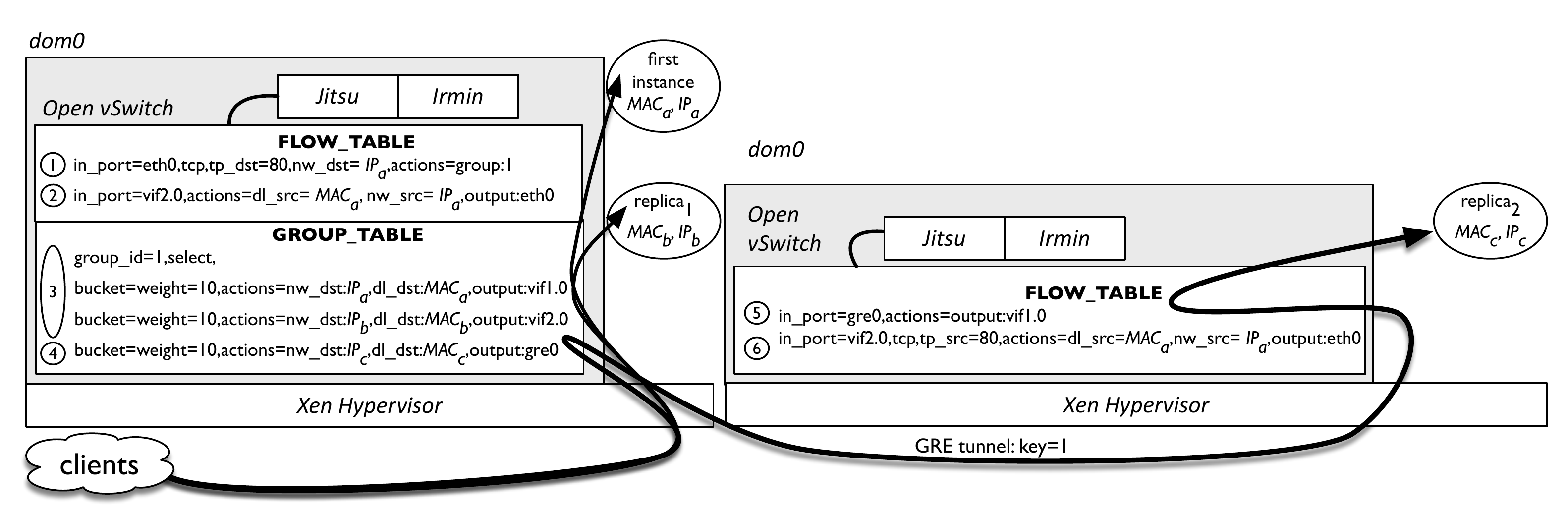}
  \caption{\label{f:of-example-policy}\system network configuration when load-balancing a web application among the first instance and two replicas. Flows are evenly distributed, after address translation, among the three instances~(\Onec, \Threec). Traffic for the remote replica is forwarded through a GRE tunnel from the first host~(\Fourc, \Fivec). Outgoing traffic from the second host has its source IP address rewritten to match the first host's~(\Twoc, \Sixc).
  \todo{SureIn figure 5, why do you special-case local calls? If Fractal scales to
      100 nodes, local calls will be rare (1\%).}}
\end{figure*}

As an example, Figure~\ref{f:of-example-policy} depicts the \of flow entries used to load balance traffic between a web application's first-instance, and two replicas, one local and one remote. Incoming traffic is direct to the  \group entry on the first host~(\Onec) while outgoing traffic is transmitted~(\Twoc). The \group selects one of three buckets uniformly randomly for each flow, and directs traffic to the relevant unikernel, after rewriting destination IP and MAC addresses~(\Threec). The situation is a little more complex for a remote replica.

To avoid the client receiving traffic from unknown sources when their requests are forwarded to a replica, we must be able to forward packets to a remote replica in compliance with datacenter policy while providing client IP and port details so that the remote replica can respond directly. We use the GRE tunnelling protocol~\cite{rfc2784} support in \ovs, configuring GRE tunnels between \ovs instances on the original and remote replica hosts through the OVSDB protocol~\cite{rfc7074}. Each GRE tunnel appears to the \ovs on the first instance's host as a port to which we forward incoming traffic via appropriate flow entries. To avoid interactions between multiple remote replicas on the same host, we create a unique tunnel for each replica with GRE keys. Our approach is not limited to the GRE tunneling protocol and could accommodate any other tunneling protocol that can integrate with an \of switch. With respect to Figure~\ref{f:of-example-policy}, rule \Fourc forwards traffic to the GRE tunnel, and flow \Fivec then forwards load balanced traffic to the remote replica.

A similar problem occurs for the outgoing traffic from remote replicas. One option would be to force all response traffic flow back via the original instance's \ovs so it could rewrite IP source address appropriately. However, this would severely limit scaling and increase cross-talk between different VMs by imposing considerable packet forwarding load on the network link and the \ovs of the original instance. Instead, we introduce controlled spoofing of source addresses by installing, in the \ovs on the replica's host, a flow that matches outgoing response packets by port number and source IP address and rewrites the source IP address to the service IP address before forwarding the packet to the NIC of the host (\Sixc in Figure~\ref{f:of-example-policy}).
\todo{Trust considerations? Why should data center allow it?
Is the fractal infrastructure trusted? That seems to oppose the design thesis.}

During development we observed that the hashing function was automatically rebalanced during updates to the \group table, resulting in delivery of packets of already established flows to the wrong replica. We remedy this using the flow caching policy of \ovs. This maintains an in-kernel flow table, inserting an equivalent exact-match flow in the kernel module when a packet is matched against a {\tt select} \group entry. We modified the {\tt ovs-vswitchd} daemon to \emph{not} remove this exact-match rule when the hash function is updated, ensuring that ongoing flows are correctly delivered during \group updates. The in-kernel cache is sized to support millions of entries, so eviction due to cache pressure is unlikely, though a proper fix would require further minor modifications to the \ovs code.

Comparing the two approaches, the second requires a minor modification to the \ovs implementation, but is standards compliant, potentially allowing offload of traffic deflection rules from the \ovs softswitch to hardware \of switches. The first, in contrast, does not require modification of the \ovs source code, but is not fully standards compliant, limiting potential future use of \of hardware switches, and also does not allow dynamic adaptation of the hashing function to the traffic load.

\begin{table}
  \centering
  \small
  \renewcommand\arraystretch{1.2}
  \begin{tabular}{l r r | r r}
    & \multicolumn{2}{c|}{\bf OVS Extension} & \multicolumn{2}{c}{\bf OF Group} \\
    \hline
    & $\mu$ & $\sigma$ & $\mu$ & $\sigma$ \\
    \hline
    & \multicolumn{4}{c}{\em CPU Utilisation}\\
    dom0     & 1.45 & 0.19 & 1.41 & 0.21 \\
    Replica A & 0.41 & 0.06 & 0.39 & 0.06 \\
    Replica B & 0.31 & 0.04 & 0.33 & 0.04 \\
    Replica C & 0.33 & 0.05 & 0.34 & 0.04 \\
    & \multicolumn{4}{c}{\em Latency (ms)}\\
    Response latency & 2.70 & 0.6 & 2.50 & 0.8 \\
    \hline
  \end{tabular}
  \caption{\label{t:ovs_comparison}Comparison between \of policies when load-balancing between first-instance, local and remote replicas of a web service. N.B.~dom0 is configured to make use of all four cores on the host, while the replicas each use only one core each due to the single-threaded OCaml runtime.}
\end{table}

Furthermore, when we evaluate the impact of each deflection policy implementation on our testbed~(\S\ref{s:evaluation}), using the topology in Figure~\ref{f:of-example-policy} to load balance a static web service, we observe that the second approach exhibits marginally better load distribution across replicas and lower (albeit by just 0.2\,ms) mean response latency. Using \httperf we generate a traffic load of 2000 page requests/second for 10 minutes, and measure the instantaneous CPU utilisation of the host's dom0 and each unikernel, as well as the response latency for each page. We configure the two network policies described above: the first uses an \of \group entry (\emph{OF Group}), and the second uses the \ovs extension discussed above (\emph{OVS Extension}). Table~\ref{t:ovs_comparison} presents the mean ($\mu$) and standard deviation ($\sigma$) of the CPU utilisation and page response latency for the duration of the experiment.

Source address spoofing should have no effect on traffic delivery as the network core of large datacenter focuses primarily on routing based on destination IPs, monitoring and link failure recovery, while finer grained traffic management policies are applied in the dom0 and the top-of-rack switches in the network.
It might raise security concerns though. However, when a replica attempts to generate malicious traffic, it can only use the IP of the first instance of the service it provides, making such attacks easily detectable and auditable. If a replica tries to spoof traffic to an arbitrary IP address, the \ovs on the destination host will drop it due to lack of a suitable flow. Any further  network management access control policies can be easily integrated by inserting appropriate higher-priority \of flows into the flow table. For example, port-based firewall policies can easily integrate with our policy by installing port-matching flows in the \path{Flow_table} with a higher priority.

\subsection{Replica Lifecycle Management}
\label{s:lifecycle}

Managing an application's use of datacenter resources requires taking decisions about when and where replicas should be created and destroyed, incorporating details about current load on datacenter components, notably servers, and network switches and links. \System makes these decisions collaboratively, with the application determining \emph{when} a replica needs to be created or destroyed, and delegating details of \emph{where} new replicas should be created to Jitsu. In turn, the datacenter operator implements policies plus any necessary monitoring infrastructure (e.g.,~load or topology monitors, or contractual details for specific customers) for Jitsu to apply to determine where replicas should be created. We do not address how to optimally place replicas in this paper, but believe that techniques used in adaptive resource management, e.g.,~\cite{eva-kalman-filters}, will be worth exploring.
\haris{VM placement is a problem for the infrastructure provider and not for the devops of an application service. \System allows the application to remain agnostic towards these decisions and enables a common control framework across the datacenter. }

The precise metrics that an application developer will wish to use depend on the application. The essence of \System is to place this decision in the hands of the developer of the specific application in question, and so we do not make any recommendations here as to which metrics should be used. The simple website example we use in our evaluation~(\S\ref{s:evaluation}) uses requests/second as its metric, with two thresholds, $(\hbox{low,high}) = (100,1000)$, determining when a unikernel decides to request creation of a new replica, or to request that itself be destroyed.
\todo{Wait, this would mean that if load is well balanced across 10 servers,
and the total load drops to 900 r/s, then all ten replicas will see about
90 r/s and shutdown simultaneously. Wouldn't central or better-coordinated
decisionmaking be in order here?}

Figure~\ref{f:irmin} depicts the metadata maintained by Jitsu for each VM. The first instance of an application VM is allocated a unique name, passed on to all replicas, and must know its unique name to write its requests under the correct key. We use the lexicographically lowest MAC address of each VM's interfaces for this purpose. Figure~\ref{f:irmin-request} shows the Irmin state involved for local and remote invocations of \emph{replicate}. The remote request must additionally supply a pathname to the Xen image to boot (assumed to be available to the remote host, e.g.,~via a network filer). Upon receiving the request, Jitsu boots the unikernel accordingly, and configures \ovs using the newly booted unikernel's domain identifier, IP address and MAC address to balance the network traffic between all instances.

For the new VM to be created on a different physical host from which the request originated, Jitsu first decides upon the target host according to whatever policies the datacenter operator has in place, and then writes the request into the target's Irmin store. Jitsu can distinguish a remote request from a local request by matching the requester's name against the names of running VMs and the list of hosts. It then defines and creates a new replica, setting the key \path{/jitsu/vm/<NAME>/initial_xs} to be the name of the requester (left blank if requester is local). The response is written back to the requesting host's Irmin. Requests are not relayed.

When a replica decides that it is no longer required, it calls \emph{halt}. In response, the local Jitsu reconfigures the first-instance \ovs to stop sending traffic to this instance. When the replica is satisfied it has finished servicing all outstanding requests, and has pushed any necessary state back to the first-instance, it calls \emph{die} which sets the TTL in its Jitsu record to zero. Jitsu has a monitoring thread that stops VMs whose TTL has reached zero, and so the replica will be shortly removed (within 5\,s by default). The \ovs on the remote machine must also be reconfigured if the replica was requested by a remote VM, indicated via the \path{initial_xs} key's value. Finally, Jitsu deletes all the Irmin keys belonging to the deleted instance.

\subsection{Failure Recovery}
\label{s:frecovery}

Failures are common in the datacenter and this section presents the strategies of \system to provide continuous service delivery. In our failure analysis we consider two failure types: \emph{replica} and \emph{host} failures. 

\smallskip
\noindent\textbf{Replica failure recovery:}
In order to detect a replica failure, \system exploits the \textit{XenStore} service, which stores realtime configuration and state information for each running VM. Effectively, during a VM failure XenStore removes all associated information from the database . \system monitors the XenStore entries and upon the removal of a replica VM information, it performs a garbage collection routine similar to a \emph{die} call. A replica reboot operation can equally lead to system inconsistencies, since Xen changes the VM DomId and the interfaces exposed to the dom0. \system can detect DomId changes for a replica, by comparing its local state with the XenStore state, triggering appropriate updates in the output action, IP and MAC address translation of the respective \ovs table flows.

\smallskip
\noindent\textbf{Xen host failure recovery:}
\System employs a heartbeat mechanism over the Irmin database to detect host failures. The Irmin database on the host of a service first instance stores a timestamp entry for each replica host. On a fixed interval, the \system instance on a replica host is responsible to update the respective entry with a timestamp, thus providing a heartbeat mechanism which allows host failure detection. On a host failure, \system removes any bucket forwarding traffic to an unavailable host. 
\todo{maybe update the layout in figure 4 and 5?}

To maintain service liveness during a failure of a first instance host we exploit the Xen High Availability (Xen HA)~\cite{xenserverHA} functionality. Xen HA enables the creation of resource pools by federating control of multiple Xen hosts, and provides automated host failure detection and VM migration. In a Xen HA pool, a host is designated as the master host, responsible for the management of VMs running on the pooled hosts. A unique instance of \system runs on the master host of the pool, managing service replication between pooled hosts, while an Irmin instance run on each host of the HA pool maintaining strong synchronisation with the master Irmin instance. Upon a non-master host failure, \system monitors any replica migration and updates accordingly the \ovs configuration of the live hosts of the pool. On a master host failure, Xen HA elects a new master host, which in turn starts an instance of \system triggered by an event hook in the Xen hypervisor. The new master \system instance will reset the networking state of the pooled hosts, reboot existing first instances and destroy all replicas. Replicated services can then use the \system API to trigger the required scale-up operations in order to meet service delivery requirements. 
\todo{do we need to discuss about network IP migration?}

\section{Evaluation}
\label{s:evaluation}

We evaluate the overheads of \System using a cluster of 4 machines with single socket 6-core Intel Xeon 2.4GHz processors and 64GB RAM to represent a rack in a datacenter. Three of the machines act as hosts in the datacenter, running XenServer 6.5; the remaining one is our traffic generator, running \httperf on Ubuntu 14.04. The prototypical scalable service our unikernel provides is a webserver serving static webpages of sizes drawn from a distribution of total transfer size for pages using statistics for September 2015 reported by HTTP Archive tracker~\cite{http-stats}. We created static pages sized to the mid-point of each bucket in those data, and fed \httperf with a seed file of 100 URLs to traverse, with URLs recurring in proportion to the frequency in the reported statistics. We use the XAPI web interface to extract network and CPU utilisation metrics.

\begin{figure}
  \centering
  \includegraphics[width=\columnwidth]{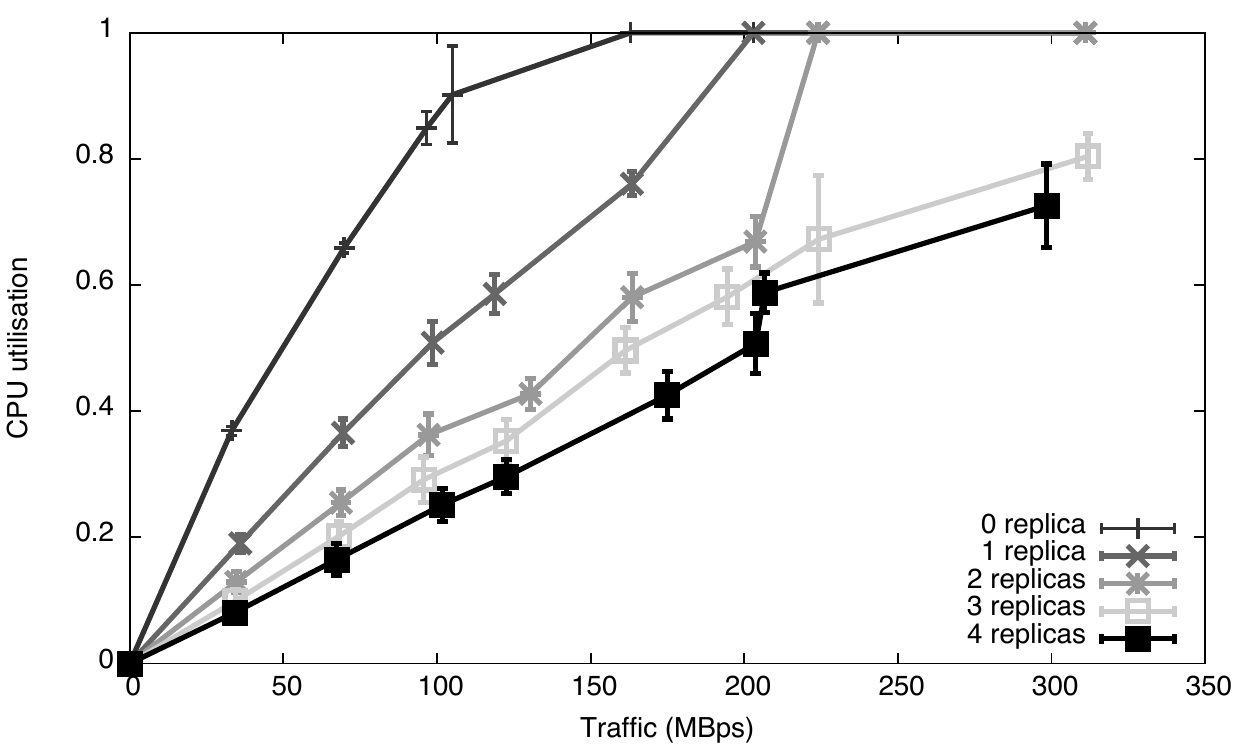}
  \caption{\label{f:exp1}Average CPU utilisation per unikernel as traffic increases linearly, for different numbers of local replicas.}
\end{figure}

{\bf Is running multiple replicas on a single host beneficial?}

The first question we address is the most basic: does our replication architecture improves the aggregate service rate. In this scenario we consider replication in a single Xen host. For each experimental run, we initialise statically \system to setup a specific replication degree and measure the CPU utilisation of each unikernel for varying traffic loads.

Figure~\ref{f:exp1} presents the CPU utilisation per unikernel for a specific traffic rate. We limit our analysis to 4 replicas due to the limited number of CPU cores (We assign two CPU core to the dom0). For each replication degree we run 10 experiments for each traffic rate and report in Figure~\ref{f:exp1} using errorbars the mean CPU utilisation and two standard deviations from the mean. From the results we highlight that  local service replication can achieve linear scalability. A non replicated service can achieve a throughput of 104~MBps, but the addition of a single replica can almost double the achievable rate (203~MBps), while the load is distributed evenly between the two replicas. However, for this configuration, it seems that once there are four instances (3 replicas and the first instance) running locally (the original instance plus three replicas), there is little benefit in creating more replicas locally. The balance between local and remote replication should clearly needs to be managed. \todo{Why is this surprising? Your host hardware has four cores,
yes?}

\begin{figure}
  \centering
  \includegraphics[width=\columnwidth]{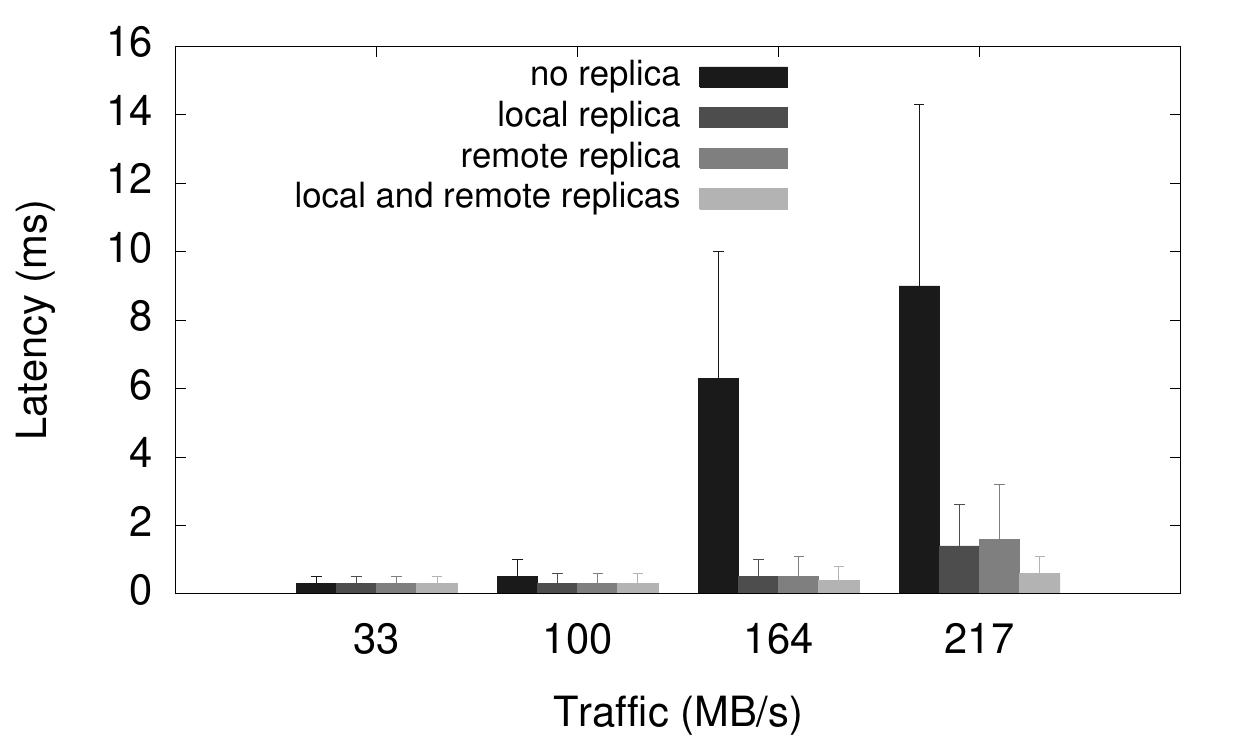}
  \caption{\label{f:exp2}Client latency experienced with no, local, and remote replicas.}
\end{figure}

{\bf How does performance compare between local and remote scaling?}

The second question we address is what is the impact on HTTP response times when a replica runs on a remote or on a local host. Figure~\ref{f:exp2} reports the mean and two standard deviation from the mean of the HTTP response latencies measured using the {\tt httperf} tool. For this measurement we employ a single web page of size 110\,kB and we vary the request rate. We measure the page download latency using four replication strategies: \emph{no replica}, which uses only the first instance of the service, \emph{local replica}, which uses a replica colocated with the first instance, \emph{remote replica}, which uses a replica located on a different host from the first instance, and \emph{local and remote replicas}, which uses a replica colocated with the first instance and a replica running on a remote host. Figure~\ref{f:exp2} presents the average page response latency for different traffic loads, with error bars indicating representing two standard errors from the mean. From the results we highlight that the \emph{no replica} strategy can provide low and predictable response latencies for traffic rates up to 100~MBps, incuring significantly inflated latencies for higher data rates. The addition of a replica achieves lower latencies up to 217~MBps~\todo{is this bytes?}, as suggested by Figure~\ref{f:exp1}. The remote replication strategy incurs a slight latency inflation, which can be explained by the required additional hop in the network path to reach the remote replica. Finally, the \emph{local and remote replicas} strategy highlights the benefit in response time for HTTP requests, when mixing local and remote replicas. \todo{Do we have a remote-remote strategy to further the discussion?}

\todo{-Unless I misunderstand, you have only 4 replicas per server because each OCaml replica is single threaded and on a 6 core server two of the cores are dedicated to dom0. Isn't this a really huge inefficiency? With more cores, would you be able to dedicated all of them to real work or do you have to scale up the dom0 piece as well.}

\begin{figure}
  \centering
  \includegraphics[width=\columnwidth]{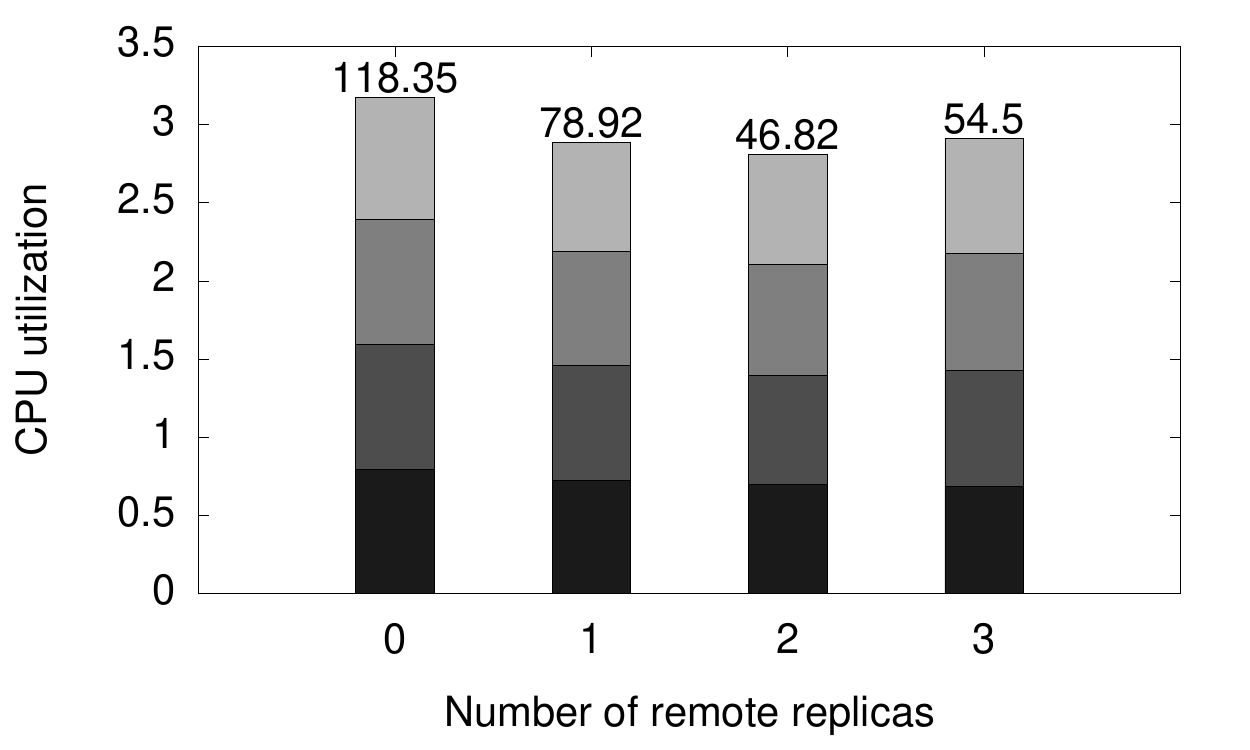}
  \caption{\label{f:exp1-lr}Average CPU utilisation per unikernel as local replicas become remote.}
\end{figure}

\begin{figure*}[t]
  \centering
  \includegraphics[width=0.9\textwidth]{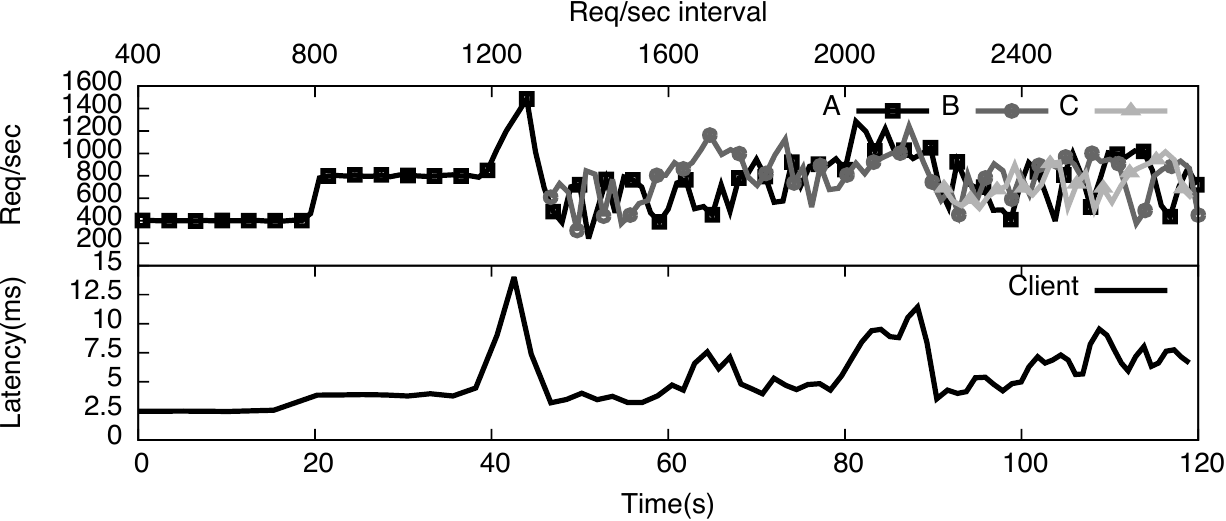}
  \caption{\label{f:exp3_req-sec}Requests per second from server side compared with client perceived latency.}
\end{figure*}

{\bf Is load distributed evenly over multiple local and remote replicas?}

Given the observed difference in the experienced latency between local and remote replication strategies, the next question we explore is whether this minor latency inflation has an impacts on the load distribution between replicas. For this experiment we setup \system to use 4 unikernels, the first instance and 3 replicas, and we differentiate the number of local and remote replicas. Figure~\ref{f:exp1-lr} presents, using a stacked bar plot, the average CPU utilisation for each unikernel under constant (high) load. We limit our analysis to 4 unikernels, because of limitations in the number of cores in the Xen host. From the results, we highlight that the aggregate CPU utilisation has small difference between the different replica strategies. We further inspected the number of packets send to each unikernel, as well as, the number of interrupts delivered to each vif and we did not find significant differences. We concluded that this marginal difference between CPU loads is observed when a high number of unikernel is hosted on a single Xen host and it is attributed to scheduling effects.

{\bf What is the impact on client response time as the service scales up and down?}

Our evaluation shows that clients are benefited during high load circumstances as depicted on Figure~\ref{f:exp3_req-sec}. This experiment consisted of sending and incrementing the traffic by 400 reqs/sec every 20 seconds using httperf, and unikernels are configured to trigger an scale up event once they hit a threshold value of 1000 reqs/sec rate. The results shows how the utility of the automated system for scaling helps alleviating the latency perceived by clients through the self-replication process. Our experiments have also shown that there is no overhead in terms of the latency perceived by the client when a unikernel that uses a self-replication mechanism but does not replicate over a unikernel that is not able to self-replicate.

\section{Related Work}
\label{s:related}

\todo{Broadly, little of the evaluation section had anything to do with the
paper's thesis. The related work section was an opportunity to hone the
paper's thesis by carefully defining the space and contrasting what makes
your approach a unique vector; I don't think this version of the related
work section achieves that goal.
}

A wide range of load-balancing services are available in commercial public cloud infrastructures. For example, Amazon AWS provides Elastic Load Balancing (ELB)~\cite{elb}, a network-based load balancing service. ELB enables cloud applications to expose multiple service replicas as a single public IP address, and dynamically load balance flows between replicated VMs. Flows are distributed via round-robin for generic TCP flows and last-connection routing for HTTP and HTTPS sessions. The load balancing service uses the Amazon VPC, a generic packet switching service used to interconnect VMs using private networks. While similar in general approach, \system hides the complexity of configuring the VPC by producing \of rules automatically, and through use of \ovs is deployable in any datacenter running Xen.

Another example on a public cloud is Azure's Ananta~\cite{Patel:2013}, a datacenter load-balancing service for the Azure datacenter. Ananta splits the load balancing function in two operations: flow distribution across service replicas and SNAT translation. Flow distribution for incoming traffic is implemented on standalone software switches, \emph{MUX}, which map each flow to a specific replica. MUX nodes integrate with the datacenter network policy through the routing protocol by advertising each service's IP reachability, and achieve scale-out capabilities via ECMP routing. SNAT functionality for outgoing traffic is implemented by software agents in the dom0 of virtualised hosts. Unlike Ananta, \system reduces the per-packet processing by integrating the per-flow mapping operation with the forwarding decision on the edge of the datacenter network, where the traffic load is lower.

Duet~\cite{gandhi:2014:sigcomm}, an extension of Ananta's design, offloads some of the replica assignment processing to hardware switches, thus improving flow latency and scalability. Duet exploits the line-rate per-packet hashing for ECMP mapping and IP-in-IP tunnelling capabilities of hardware switches, which remain underutilised in modern datacenters. Nonetheless, the respective lookup tables are limited and cannot store global load balancing state. Duet implements a routing policy which localise traffic of a load balanced IP in a single switch and defines a distribution algorithm that maximize service coverage. Software switches are used for fail-over redundancy and traffic processing when the mapping state for a load balanced IP does not fit in the hardware switch lookup tables. 

Application controlled management has been explored in the context of virtualized environments. The Potemkin honeypot farm~\cite{Vrable:2005:sosp} was a highly available honeypot farm that could spawn new honeypot VMs during an attack and was an early attempt ro define an application API for VM management. The authors developed a set of extensions in the Click~\cite{click} router that triggered the creation of dedicated honeypots and forwarded traffic to it when an attack was positively detected; during idle periods, those VMs would be garbage collected. Another example is SnowFlock~\cite{Lagar-Cavilla:2009:eurosys} which provides a Xen-specific \emph{pthread}-inspired fork API enabling rapid parallel VM migration using network multicast for the distribution of the OS state. While the API is superficially similar, the authors primarily focused on the optimisation of the migration process and did not discuss integration with the complex network policies used in modern multi-tenant datacenters.

Datacenter research has also explored mechanisms to automate cloud application replication. AGILE~\cite{agile}, a notable service replication automation framework, uses wavelets analysis to predict application resource requirements in order to fulfil specific service delivery guarantees and proactively optimize replication strategies. AGILE define a generic black-box performance modeling framework, which uses monitoring agents to collect low-level information, like CPU, memory and network utlization for each VM and predicts future resource requirements. Unlike AGILE, \system enables applications to define and run application centric scaling mechanisms with greater agility.   



Several commercial providers offer products with hardware-accelerated load balancing features, e.g.,~A10networks~\cite{a10networks}, Brocade~\cite{brocade} and Baracuda~\cite{baracuda}, though these have high deployment costs and require considerable effort to integrate them with datacenter network policies. Widespread adoption of IaaS services has even motivated development of dedicated load balancing cloud appliances. For example, the VA R20 virtual appliance~\cite{loadbalancer} is a VM image which provides flexible layer-4 to layer-7 load balancing between replicated VMs. Similarly, HAProxy~\cite{haproxy} is a popular open-source application used to load balance cloud applications using dedicated Linux VMs. Such cloud appliances make deployment straightforward for self-managed load-balanced services, but exhibit limited scale up capabilities and inflate per-packet latency and resource utilisation, as each client request must traverse at least two VMs. Cloud management platforms, such as the OpenStack project~\cite{openstack}, also offer load balancing service management capabilities. OpenStack has developed a generalizable load-balancing control abstraction through the Neutron LBaaS project~\cite{neutron-lbaas}, and provides integration with a wide range of software and hardware platforms. Similarly, the Kubernetes~\cite{kubernetes} container management platform provide a framework to control the service replication strategy for a service. Specifically, it employs per-cluster replication controllers which execute application centric health checks, like HTTP status code monitoring, network connectiivity, and command executability within the container, in order to maintain a constant replication degree. The Kubernetes replication controller provides a composable primitive which allows scaling and scheduling automation from external controllers through a simple API. For example, Kubernetes provides a beta horizontal autoscaler functionality which allows a users to define a targer average CPU utilization during service deployment and the autoscaler will ensure to meet this target by appropriately controlling the replication degree of the service. \System proposes an application-centric orchestration , which integrates the application and the orchestration logic.

Load balancing was one of the early application use cases for programmable network control and a number of early network control architectures~\cite{Wang:2011,handigol2009plug} have explored reactive and proactive \of-based schemes. These architectures, primarily explored the applicability of control plane programmability to achieve load-balancing functionality, but did not explore in-depth an holistic SDN solution. Furthermore, the introduction of the SDN paradigm has motivated the design of novel load balancing control abstractions. Mekky~\etal~\cite{mekky:2014:hotsdn} proposed an enhancement in the \of protocol for application-specific extensions in the packet processing pipeline, which enables integration of arbitrary packet processing applications in the data plane. Pane~\cite{ferguson:2012:HotICE} is a control architecture which enables control delegation to end-user applications. The framework provides a generalisable mechanism to synthesise individual policies in a single network configuration, while policy conflicts are resolved by direct negotiation with the requesting applications. \system exposes a control abstraction focused on load balancing service control, while policy conflicts can be resolved by appropriate prioritisation of flows. Finally, Mosyre~\etal~\cite{NiranjanMysore:2013:connext} quantify the VM network performance limitations in 10 GbE datacenters and explore the effectiveness of multiple network operation offloading mechanisms in datacenters (SR-IOV VF, ToR tunnelling offloading) to improve throughput. We plan to explore similar offloading opportunities to further improve the scalability of our system.

\section{Conclusions}
\label{s:conclusions}

In conclusion, modern cloud service design is heavily dominated by the paradigm of separation between application logic and deployment configuration and management. This approach limits the ability to develop flexible mechanisms that dynamically adapt the service configuration in order to meet application-specific demands and service requirements. This paper presents \system, an alternative approach to cloud service management which allows embedding of service management in the application logic of the service.

\System extends the Jitsu control stack on the dom0 through a simple RPC API which allows simple, dynamic, programmable creation and destruction of service replicas over the cloud infrastructure from within the service. This enables such decisions to be taken at whatever granularity and based on whatever metrics the developer sees fit. \system implements network-based load balancing using the \of protocol as exposed by the \ovs softswitch running in the dom0 of the XenServer. As a proof of concept of the applicability of \system, we have developed a simple self-scaling web service based on a MirageOS webserver. Evaluating this \system-enabled unikernel shows that the proposed system does not have significant negative performance impact, while effectively providing flexible load distribution across replicas.

Although we have presented and evaluated \system in the context of MirageOS unikernels, it is not limited to either MirageOS or unikernels but can be adopted by other applications. For example, both \emph{nginx} and Apache define an extensive plugin API which can extract important application measurements that could be used by scripts to invoke \system API. However, the introduction of VMs with disks, storage and other less straightforwardly replicated components could affect both the performance and the complexity of the replication and destruction processes. Recent developments using \emph{rumpkernels} to convert legacy applications into unikernels mitigate these problems however, and may provide an excellent middle ground~\cite{cacm-unik}.

The use of Irmin to store application state potentially offers more significant benefits than we have demonstrated here. By taking what would be kernel datastructures in a traditional OS environment, and managing them inside Irmin in the unikernel, it becomes possible to replicate such state. For example, this might enable the state of a running unikernel's ARP stack to be replicated along with the unikernel itself if the replica was created within the same subnet, reducing both the time it takes for the replica to become active  and the potential for ARP storms~\cite{irminarp}.

In summary, we believe that \system provides the basis for a useful and novel service management framework that matches the dynamic nature, short timescale variability and complex configuration needs of modern microservice architectures. Although most powerful when configured for use with modern unikernels and a storage system such as Irmin, \system is also applicable in more traditional datacenter contexts. \system is available under a BSD licence at \url{https://github.com/koleini/jitsu}.

\begin{acks}
  This work was funded in part by UK EPSRC EP/K031724/2, EP/G065802/1, and EP/M02315X/1.
\end{acks}



{
  \bibliographystyle{abbrv}
  \bibliography{ms}

\begin{thebibliography}{10}

\bibitem{xenserverHA}
High availability for {Citrix XenServer}.
\newblock \url{www.citrix.com}.

\bibitem{a10networks}
{A10networks thunder ADC}.
\newblock
  \url{https://www.a10networks.com/sites/default/files/A10-DS-15100-EN.pdf}.

\bibitem{elb}
{AWS Elastic Load Balancing}.
\newblock \url{https://aws.amazon.com/elasticloadbalancing/}.

\bibitem{ansible}
Ansible inc.
\newblock \url{http://www.ansible.com}.

\bibitem{baracuda}
{Barracuda Load Balancer ADC}.
\newblock
  \url{https://www.barracuda.com/assets/docs/dms/Barracuda_Load_Balancer_ADC_DS_US.pdf}.

\bibitem{xen}
P.~Barham, B.~Dragovic, K.~Fraser, S.~Hand, T.~Harris, A.~Ho, R.~Neugebauer,
  I.~Pratt, and A.~Warfield.
\newblock {Xen and the Art of Virtualization}.
\newblock In {\em Usenix SOSP}, 2003.

\bibitem{kubernetes}
E.~Brewer.
\newblock Kubernetes and the path to cloud native.
\newblock In {\em ACM SoCC}, 2015.
\newblock Keynote.

\bibitem{brocade}
{Brocade ADX Series}.
\newblock
  \url{http://www.brocade.com/en/products-services/application-delivery-controllers/adx-series.html}.

\bibitem{chef}
Chef software inc.
\newblock \url{https://www.chef.io/}.

\bibitem{irmin}
B.~Farinier, T.~Gazagnaire, and A.~Madhavapeddy.
\newblock {Mergeable persistent data structures}.
\newblock In {\em {JFLA 2015}}.

\bibitem{ferguson:2012:HotICE}
A.~D. Ferguson, A.~Guha, J.~Place, R.~Fonseca, and S.~Krishnamurthi.
\newblock Participatory networking.
\newblock In {\em {Usenix HotICE}}, 2012.

\bibitem{gandhi:2014:sigcomm}
R.~Gandhi, H.~H. Liu, Y.~C. Hu, G.~Lu, J.~Padhye, L.~Yuan, and M.~Zhang.
\newblock Duet: Cloud scale load balancing with hardware and software.
\newblock In {\em ACM SIGCOMM}, 2014.

\bibitem{oxenstored}
T.~Gazagnaire and V.~Hanquez.
\newblock {OXenstored}: an efficient hierarchical and transactional database
  using functional programming with reference cell comparisons.
\newblock {\em SIGPLAN Notices}, 44(9), 2009.

\bibitem{handigol2009plug}
N.~Handigol, S.~Seetharaman, M.~Flajslik, N.~McKeown, and R.~Johari.
\newblock Plug-n-serve: Load-balancing web traffic using openflow (demo).
\newblock 2009.

\bibitem{rfc2784}
S.~Hanks, D.~Meyer, D.~Farinacci, and P.~Traina.
\newblock Generic routing encapsulation {(GRE)}.
\newblock \url{https://tools.ietf.org/html/rfc2784}, Mar. 2000.

\bibitem{haproxy}
{HAProxy}.
\newblock \url{http://www.haproxy.org/}.

\bibitem{http-stats}
{HTTP Archive Interesting Stats}.
\newblock \url{http://www.httparchive.org/interesting.php}, Sept. 2015.

\bibitem{autopilot}
M.~Isard.
\newblock Autopilot: Automatic data center management.
\newblock {\em SIGOPS Oper. Syst. Rev.}, 41(2), 2007.

\bibitem{eva-kalman-filters}
E.~Kalyvianaki, T.~Charalambous, and S.~Hand.
\newblock Adaptive resource provisioning for virtualized servers using kalman
  filters.
\newblock {\em ACM Trans. Auton. Adapt. Syst.}, 9(2), 2014.

\bibitem{click}
E.~Kohler, R.~Morris, B.~Chen, J.~Jannotti, and M.~F. Kaashoek.
\newblock The click modular router.
\newblock {\em ACM Trans. Comput. Syst.}, 18(3), 2000.

\bibitem{Lagar-Cavilla:2009:eurosys}
H.~A. Lagar-Cavilla, J.~A. Whitney, A.~M. Scannell, P.~Patchin, S.~M. Rumble,
  E.~de~Lara, M.~Brudno, and M.~Satyanarayanan.
\newblock Snowflock: Rapid virtual machine cloning for cloud computing.
\newblock In {\em ACM EuroSys}, 2009.

\bibitem{loadbalancer}
loadbalancer cloud products.
\newblock \url{http://www.loadbalancer.org/uk/products/cloud}.

\bibitem{jitsu}
A.~Madhavapeddy, T.~Leonard, M.~Skjegstad, T.~Gazagnaire, D.~Sheets, D.~Scott,
  R.~Mortier, A.~Chaudhry, B.~Singh, J.~Ludlam, J.~Crowcroft, and I.~Leslie.
\newblock Jitsu: Just-in-time summoning of unikernels.
\newblock In {\em USENIX NSDI}, 2015.

\bibitem{mirage}
A.~Madhavapeddy, R.~Mortier, C.~Rotsos, D.~Scott, B.~Singh, T.~Gazagnaire,
  S.~Smith, S.~Hand, and J.~Crowcroft.
\newblock Unikernels: Library operating systems for the cloud.
\newblock In {\em ACM ASPLOS}, 2013.

\bibitem{cacm-unik}
A.~Madhavapeddy and D.~J. Scott.
\newblock Unikernels: The rise of the virtual library operating system.
\newblock {\em Commun. ACM}, 57(1):61--69, Jan. 2014.

\bibitem{mekky:2014:hotsdn}
H.~Mekky, F.~Hao, S.~Mukherjee, Z.-L. Zhang, and T.~V. Lakshman.
\newblock Application-aware data plane processing in {SDN}.
\newblock In {\em {ACM HotSDN}}, 2014.

\bibitem{mesos}
Apache mesos.
\newblock \url{https://mesos.apache.org/}.

\bibitem{agile}
H.~Nguyen, Z.~Shen, X.~Gu, S.~Subbiah, and J.~Wilkes.
\newblock {AGILE}: elastic distributed resource scaling for
  infrastructure-as-a-service.
\newblock In {\em {USENIX ICAC}}, 2013.

\bibitem{NiranjanMysore:2013:connext}
R.~Niranjan~Mysore, G.~Porter, and A.~Vahdat.
\newblock {FasTrak}: Enabling express lanes in multi-tenant data centers.
\newblock In {\em ACM CoNEXT}, 2013.

\bibitem{neutron-lbaas}
Neutron load balance as a service.
\newblock \url{https://github.com/openstack/neutron-lbaas}.

\bibitem{openstack}
{OpenStack}.
\newblock \url{http://www.openstack.org/}.

\bibitem{Patel:2013}
P.~Patel, D.~Bansal, L.~Yuan, A.~Murthy, A.~Greenberg, D.~A. Maltz, R.~Kern,
  H.~Kumar, M.~Zikos, H.~Wu, C.~Kim, and N.~Karri.
\newblock Ananta: Cloud scale load balancing.
\newblock In {\em ACM SIGCOMM}, 2013.

\bibitem{rfc7074}
B.~Pfaff and B.~Davie.
\newblock {The Open vSwitch Database Management Protocol}.
\newblock \url{https://tools.ietf.org/html/rfc7047}, Dec. 2013.

\bibitem{ovs}
B.~Pfaff, J.~Pettit, T.~Koponen, E.~Jackson, A.~Zhou, J.~Rajahalme, J.~Gross,
  A.~Wang, J.~Stringer, P.~Shelar, K.~Amidon, and M.~Casado.
\newblock The design and implementation of open vswitch.
\newblock In {\em USENIX NSDI}, 2015.

\bibitem{irminarp}
M.~Preston, M.~Skjegstad, T.~Gazagnaire, R.~Mortier, and A.~Madhavapeddy.
\newblock Persistent networking with irmin and {MirageOS}.
\newblock In {\em Proc. OCaml Workshop at ICFP}, Sept. 2015.

\bibitem{puppet}
Puppet labs.
\newblock \url{https://puppetlabs.com/}.

\bibitem{borg}
A.~Verma, L.~Pedrosa, M.~Korupolu, D.~Oppenheimer, E.~Tune, and J.~Wilkes.
\newblock {Large-scale Cluster Management at Google with Borg}.
\newblock In {\em {ACM EuroSys}}, 2015.

\bibitem{Vrable:2005:sosp}
M.~Vrable, J.~Ma, J.~Chen, D.~Moore, E.~Vandekieft, A.~C. Snoeren, G.~M.
  Voelker, and S.~Savage.
\newblock Scalability, fidelity, and containment in the potemkin virtual
  honeyfarm.
\newblock {\em SIGOPS Oper. Syst. Rev.}, 39(5), 2005.

\bibitem{Wang:2011}
R.~Wang, D.~Butnariu, and J.~Rexford.
\newblock Openflow-based server load balancing gone wild.
\newblock In {\em USENIX HotICE}, 2011.

\end{thebibliography}
}
\end{document}